\documentclass[9pt,twocolumn,twoside]{opticajnl}
\journal{opticajournal} % use for journal or Optica Open submissions

\setboolean{shortarticle}{false}
\usepackage{lineno}
\usepackage{upgreek}
\usepackage{amsfonts}
\usepackage{mathrsfs}
\usepackage{wasysym}
\usepackage{amsmath}
\usepackage{amssymb}
\usepackage{color}
\usepackage{graphicx}
\usepackage{xspace}
\usepackage{enumerate}
\usepackage{tikz}
\usepackage{pstricks}
\usepackage{hyperref}
\usepackage{dsfont}
\usepackage{pifont}
\usepackage{multirow}
\usepackage{balance}
\usepackage{hyperref}
\hypersetup{
	colorlinks=true,
	linkcolor=blue,
	filecolor=blue,      
	urlcolor=blue,
	citecolor=blue
}
\usepackage{booktabs}
\usepackage{bm}
\usepackage{palatino}
\usepackage{array,graphicx}

\title{Topological edge states in photonic Floquet insulator with unpaired Dirac cones}

\author[1]{Hua Zhong} 
\author[2]{Yaroslav V. Kartashov} 
\author[1]{Yongdong Li} 
\author[3]{Ming Li}
\author[1,*]{Yiqi Zhang}

\affil[1]{Key Laboratory for Physical Electronics and Devices, Ministry of Education, School of Electronic Science and Engineering, Xi'an Jiaotong University, Xi'an 710049, China}
\affil[2]{Institute of Spectroscopy, Russian Academy of Sciences, Troitsk, Moscow, 108840,	Russia}
\affil[3]{State Key Laboratory of Transient Optics and Photonics, Xi'an Institute of Optics and Precision Mechanics of Chinese Academy of Sciences, 710119 Xi'an, China}

\affil[*]{zhangyiqi@xjtu.edu.cn}

\begin{abstract}
	Topological insulators are most frequently constructed using lattices with specific degeneracies in their linear spectra, such as Dirac points. For a broad class of lattices, such as honeycomb ones, these points and associated Dirac cones generally appear in non-equivalent pairs. Simultaneous breakup of the time-reversal and inversion symmetry in systems based on such lattices may result in the formation of the unpaired Dirac cones in bulk spectrum, but the existence of topologically protected edge states in such structures remains an open problem. Here photonic Floquet insulator on honeycomb lattice with unpaired Dirac cones in its spectrum is introduced that can support unidirectional edge states appearing at the edge between two regions with opposite sublattice detuning. Topological properties of this system are characterized by the nonzero valley Chern number. Remarkably, edge states in this system can circumvent sharp corners without inter-valley scattering even though there is no total forbidden gap in the spectrum. Our results reveal unusual interplay between two different physical mechanisms of creation of topological edge states based on simultaneous breakup of different symmetries of the system.
\end{abstract}

\setboolean{displaycopyright}{false} % Do not include copyright or licensing information in submission.

\begin{document}
	
	\maketitle
	
	\section{Introduction}

	Dirac points are specific degeneracies, where two bands touch each other, representative for linear spectra of many periodic systems, including honeycomb, triangular, kagome, and Lieb lattices. Dispersion relation is linear around Dirac points that leads to unusual dispersion dynamics of the wavepackets exciting these points~\cite{leykam.aipx.1.101.2016}. A variety of physical effects associated with the presence of Dirac points are reported in material sciences, acoustics, optics and physics of matter waves. In particular, honeycomb lattices with Dirac degeneracies~\cite{peleg.prl.98.103901.2007, song.nc.6.6272.2015, zhang.nc.11.1902.2020, yan.aom.8.1902174.2020}, representing photonic analogues of graphene~\cite{neto.rmp.81.109.2009}, are widely used for construction of topological insulators -- novel materials, where excitations can propagate along the edges of the structure, but not in its bulk~\cite{lu.np.8.821.2014, ozawa.rmp.91.015006.2019, smirnova.apr.7.021306.2020, yan.aom.2001739.2021}. In usual honeycomb lattices Dirac points emerge in six corners of the Brillouin zone, in pairs corresponding to non-equivalent $\textbf{K}$ and $\textbf{K}'$ points. However, in \cite{haldane.prl.61.2015.1988} a system with ``parity anomaly'' was suggested, which may possess unpaired Dirac cones if the time-reversal symmetry and the inversion symmetry of the system are broken simultaneously. The existence of the unpaired Dirac cones implies that massless chiral fermions in such a system loose anomaly-cancelling partners with opposite chirality, and thus their appearance breaks the ``fermion doubling'' theorem~\cite{haldane.prl.61.2015.1988}. However, due to unique nature of the unpaired Dirac points and specific symmetry properties that the system possessing them should manifest, the search of such materials is a challenging and open problem.
	
	Different approaches can be used to break time-reversal symmetry in photonic or optoelectronic systems. Among them is the utilization of gyromagnetic optical materials in external magnetic fields~\cite{wang.nature.461.772.2009}, or polariton condensates with pronounced spin-orbit coupling in magnetic fields~\cite{nalitov.prl.114.116401.2015, kartashov.optica.3.1228.2016, klembt.nature.562.552.2018}, see also review~\cite{ozawa.rmp.91.015006.2019}. This effect can also be achieved by using longitudinal modulations in waveguide-based systems, such as photonic Floquet insulators on helical waveguide arrays~\cite{rechtsman.nature.496.196.2013}, where a number of topological phenomena have been predicted and observed in linear and nonlinear regimes~\cite{lumer.prl.111.243905.2013, ablowitz.pra.96.043868.2017, stuetzer.nature.560.461.2018, lustig.nature.567.356.2019}. In these latter systems the breakup of time-reversal symmetry is achieved due to the effective gauge field arising from waveguide modulations~\cite{rudner.prx.3.031005.2013,rudner.nrp.2.229.2020,yin.elight.2.8.2022}.
	By the way, the Floquet waveguides arrays always provide promising results in light manipulations~\cite{song.ap.2.036001.2020,xu.ap.5036005.2023,zhang.light.13.99.2024}. 
	Unpaired Dirac points can, in principle, be observed in helical \textit{square} waveguide arrays, as suggested in~\cite{leykam.prl.117.013902.2016,noh.np.13.611.2017}, where introduction of difference into phases of helical waveguide rotation in two sublattices also breaks inversion symmetry of the structure~\cite{liang.prl.110.203904.2013, pasek.prb.89.075113.2014, chong.jo.18.014001.2016}. Experimentally, unpaired Dirac points were observed for microwaves in gyromagnetic materials~\cite{liu.nc.1873.2020}, while transitions between regimes with broken inversion and time-reversal symmetries in gyromagnetic structures were reported in~\cite{wang.lpor.18.2300764.2024}. 
	Very recently, unpaired Dirac points were also proposed and observed in a photonic crystal consisting of Y-shaped gyromagnetic rods~\cite{wang.nc.14.4457.2023}.
	Nevertheless, unpaired Dirac points remain elusive at optical frequencies in waveguiding systems. Topological edge states, that should have unique properties in such materials connected with unusual structure of their linear spectra lacking total gap, have never been reported, to the best of our knowledge.
	
	In this work, we show that such topological edge states can be observed when one simultaneously introduces into a \textit{honeycomb} waveguide array the helicity of waveguides, resulting in breakup of its time-reversal symmetry, and also the detuning between its two sublattices, resulting in breakup of the inversion symmetry of the structure. Notice that the approach utilizing sublattice detuning~\cite{noh.prl.120.063902.2018, wu.nc.8.1304.2017, zhong.ap.3.056001.2021, tang.oe.29.39755.2021, ren.nano.10.3559.2021, tian.fop.17.53503.2022, tang.chaos.161.112364.2022} is also quite involved in the investigations of the valley Hall effect~\cite{xiao.rmp.82.1959.2010} where, however, the system possesses the total gap, in contrast to system with unpaired Dirac points considered here. We show that the bulk spectrum of our system has unique structure with gap opened between, say, three $\textbf{K}'$ points, and preserved Dirac cones in non-equivalent to them three $\textbf{K}$ points. Remarkably, we find that the domain wall between two arrays with different signs of sublattice detuning supports topologically protected edge states, which, in complete contrast to valley Hall systems, feature asymmetric projected dependence of quasi-propagation constant on Bloch momentum and connect two different bands. Moreover, such edge states are topologically protected and can circumvent sharp corners of the domain wall without backscattering, despite the absence of complete gap in the spectrum of this system. Our results suggest a new platform for observation of the unpaired Dirac cones and open new prospects for investigation of nonlinear effects in topological systems lacking complete gap.
	
	\section{Theoretical model}
	
	\subsection{Helical waveguide array and its band structure}
\begin{figure}[htpb]
		\centering
		\includegraphics[width=\columnwidth]{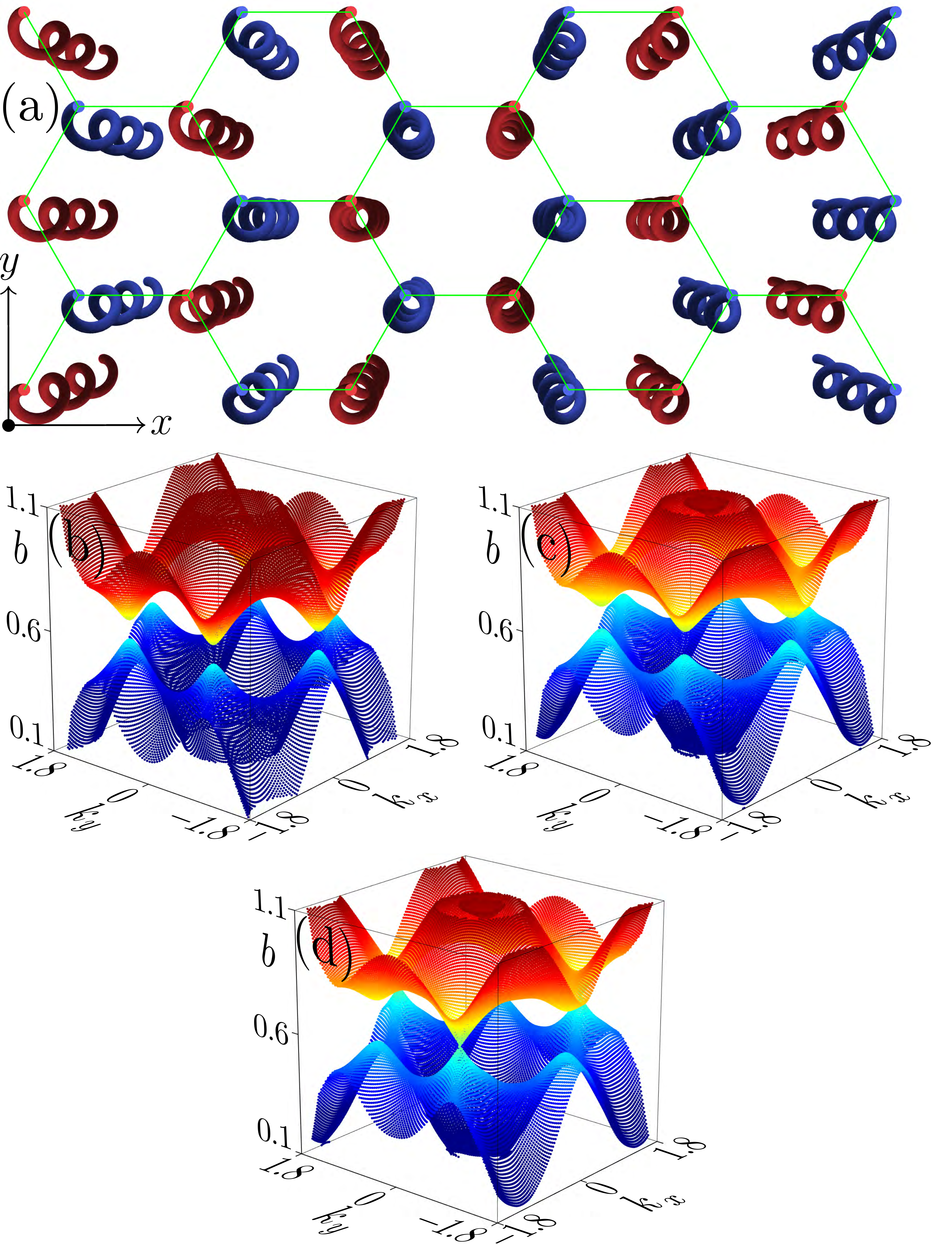}
		\caption{(a) Schematic illustration of helical waveguide array with detuning between two sublattices, shown with red and blue colors. Green hexagons highlight honeycomb structure of the array. Band structure of bulk waveguide array with ${r=0}$ and ${\delta=0.1}$ (b), ${r=0.4}$ and ${\delta=0}$ (c), and ${r=0.4}$ and ${\delta=0.1}$ (d).}
		\label{fig1}
	\end{figure}

	The propagation dynamics of a laser beam in helical shallow waveguide array can be described by the Schr\"odinger-like equation for normalized amplitude $\psi$ of the light field:
	\begin{equation}\label{eq1}
		i \frac{\partial \psi}{\partial z}=-\frac{1}{2}\left(\frac{\partial^2}{\partial x^2}+\frac{\partial^2}{\partial y^2}\right) \psi-\mathcal{R}(x, y, z) \psi,
	\end{equation}
	where $x,\,y$ are the transverse coordinates which are to the characteristic scale ${r_0=10\,\mu\textrm{m}}$, 
	$z$ is the propagation distance that is normalized to the diffraction length ${kr_0^2\approx1.14\,\textrm{mm}}$,
		${k=2\pi n/\lambda}$ is the wavenumber in the medium with the background refractive index $n$ (for fused silica ${n\approx 1.45}$), 
		and ${\lambda=800~\textrm{nm}}$ is the working wavelength.
	$\mathcal{R}(x,y,z)$ describes waveguide array composed from helical Gaussian-shaped waveguides
	\begin{equation}\label{eq2}
		\mathcal{R}=\sum_{m,n}p_{m,n}\exp\left(-\frac{[x-x_{m,n}(z)]^2+[y-y_{m,n}(z)]^2}{d^2}\right).
	\end{equation}
	Here, $d$ is the waveguide width, $(x_{m,n},y_{m,n})$ are the $z$-dependent coordinates of waveguide centers in honeycomb grid, ${p_{m,n}=k^2r_0^2\delta n/n}$ is the waveguide depth that is proportional to the refractive index contrast $\delta n$ in the array (for instance, $p=1.0$ corresponds to ${\delta n\sim 1.1\times10^{-4}}$), spacing between neighboring waveguides in the grid is equal to $a$. In helical waveguide array coordinates of waveguides change with $z$ as
	\begin{equation*}
		\left\{
		\begin{split}
			&{x_{m,n}(z) = x_{m,n}(0)+r\sin(\Omega z)}, \\
			&{y_{m,n}(z) = y_{m,n}(0)+r\cos(\Omega z)-r},
		\end{split}
		\right.
	\end{equation*}
	where $r$ is the radius of helix and ${\Omega=2\pi/Z}$ is the rotation frequency defined by period $Z$. We further introduce detuning between two sublattices of the array, by setting depth as ${p_{m,n}=p+\delta}$ or ${p_{m,n}=p-\delta}$, where $\delta$ is the depth detuning that is connected with different refractive index contrast in two sublattices and that corresponds to different on-site ``energies'' in the language of condensed-matter physics~\cite{noh.prl.120.063902.2018,zhong.adp.529.1600258.2017,tang.oe.29.39755.2021}, as shown in schematic array illustration in Fig.~\ref{fig1}(a). Further we set the mean waveguide depth ${p=8.9}$, typical value of detuning ${\delta=0.1}$, waveguide spacing ${a=1.6}$ (corresponding to $16\,\mu\rm m$), width ${d=0.4}$ (corresponding to $4\,\mu\rm m$), helix period ${Z=6}$ (corresponding to $6.8\,\rm mm$), and radius ${r=0.4}$ (corresponding to $4\,\mu\rm m$). These parameters are typical for waveguide arrays inscribed in fused silica using focused pulses from femtosecond laser~\cite{rechtsman.nature.496.196.2013, kirsch.np.17.995.2021, ren.light.12.194.2023, arkhipova.sb.68.2017.2023,tan.ap.3.024002.2021,li.ap.4.024002.2022}.
	
	\begin{table}[htbp]
		\centering
		\caption{Sign of the Berry curvature $\mathcal{B}$ of valleys for different signs of detuning and different directions of waveguide rotation (helicity). ${X}$ means that the Dirac cone may appear.
		The rotation direction (clockwise $\circlearrowright$ or anti-clockwise $\circlearrowleft$) of the waveguide array is defined along negative $z$ direction (viz. counter to the light propagation direction). }
		\label{table1}
		\setlength{\tabcolsep}{7.5mm}{
			\begin{tabular}{*{9}{c}} \toprule
				&                      & $\mathcal{B}({\bf K})$ &  $\mathcal{B}({\bf K}')$ \\ \hline
				\multirow{3}{*}{\ding{172}} &  ${\delta>0}$        & $-$                    & $+$                      \\ \cline{2-4}
				&  $\circlearrowright$ & $+$                    & $+$                      \\ \cline{2-4}
				&  total               & $X$           & $+$                      \\ \midrule
				\multirow{3}{*}{\ding{173}} &  ${\delta>0}$        & $-$                    & $+$                      \\ \cline{2-4}
				&  $\circlearrowleft$  & $-$                    & $-$                      \\ \cline{2-4}
				&  total               & $-$                    & $X$             \\ \midrule
				\multirow{3}{*}{\ding{174}} &  ${\delta<0}$        & $+$                    & $-$                      \\ \cline{2-4}
				&  $\circlearrowright$ & $+$                    & $+$                      \\ \cline{2-4}
				&  total               & $+$                    & $X$             \\ \midrule
				\multirow{3}{*}{\ding{175}} &  ${\delta<0}$        & $+$                    & $-$                      \\ \cline{2-4}
				&  $\circlearrowleft$  & $-$                    & $-$                      \\ \cline{2-4}
				&  total               & $X$           & $-$                      \\ \bottomrule
		\end{tabular}}
	\end{table}
	
	To analyse the impact of helicity and detuning on band structure, we first consider Floquet-Bloch modes of bulk array that can be found as ${\psi(\bm{r},z)=\phi(\bm{r},z)\exp(ibz+i\bm{k}\cdot\bm{r})}$, where the Floquet-Bloch function $\phi(\bm{r},z)$ is transversely and $Z$-periodic, $b$ is the quasi-propagation constant, $\bm{k}=(k_x,k_y)$ is the Bloch momentum. The function $\phi$ and the dependence $b({\bm k})$ can be found by solving the equation
	\begin{equation}\label{eq3}
		b \phi=\frac{1}{2}\left(\frac{\partial^2}{\partial x^2}+\frac{\partial^2}{\partial y^2}\right) \phi+\mathcal{R}(x, y, z) \phi+i \frac{\partial \phi}{\partial z}
	\end{equation}
	using the ``propagation and projection'' method~\cite{leykam.prl.117.013902.2016, ren.chaos.166.113010.2023, shen.prap.20.014012.2023}, which combines plane-wave expansion for calculation of eigenmodes of static array and their subsequent propagation in helical structure for calculation of quasi-propagation constants and Floquet-Bloch modes. The spectrum of static honeycomb array at $\delta, r=0$ is known to posses six Dirac points, belonging to two groups $\textbf{K}$ and $\textbf{K}'$, where top and bottom bands touch. If only the detuning $\delta$ is introduced into structure to break its inversion symmetry, while the waveguides are straight $(r=0)$, all Dirac cones split and become valleys, and complete gap opens in the spectrum of the system [Fig.~\ref{fig1}(b)]. Similar transformation of the spectrum is observed for zero detuning $\delta=0$, when waveguides are made helical, that leads to the breakup of the time-reversal symmetry of the system [see spectrum in Fig.~\ref{fig1}(c)]. It should be stressed, however, that even though Dirac cones split in both Figs.~\ref{fig1}(b) and \ref{fig1}(c), the Berry curvature (see definition in Section~\ref{topo}) in valleys behave differently: For upper band in Fig.~\ref{fig1}(c) the curvature becomes positive around all $\textbf{K}$ and $\textbf{K}'$ points, while in Fig.~\ref{fig1}(b) it acquires opposite signs in $\textbf{K}$ and $\textbf{K}'$ points (Berry curvature in the lower band in a given valley is usually opposite to that in the upper band). Therefore, simultaneous breakup of time-reversal symmetry and inversion symmetry in system with $\delta, r \ne 0$ leads to different variations of Berry curvature in $\textbf{K}$ and $\textbf{K}'$ points, that may translate into competing impact of helicity and detuning on gap width around these points (as originally predicted in Haldane model \cite{haldane.prl.61.2015.1988} for electronic system). Thus, we find that at ${r=0.4}$ the increase of sublattice detuning up to ${\delta\approx0.10}$ (for ${r_0=0.3}$ up to ${\delta\approx0.07}$), leads to restoration of the Dirac cones in three $\textbf{K}$ points, while gap remains open in three $\textbf{K}'$ points, as shown in Fig.~\ref{fig1}(d), i.e. unpaired Dirac cones appear in the bulk spectrum of this system. By changing sign of detuning $\delta$, one achieves opening of the gap in $\textbf{K}$ points instead, while Dirac points are observed in $\textbf{K}'$ points in this case.
	
	Possible scenarios of the band structure transformation are summarized in Table~\ref{table1}, where we use the fact that breakup of the inversion symmetry tends to create valleys, where the Berry curvature in each band is an odd function of $\bm k$, while breakup of time-reversal symmetry tends to create valleys, where the Berry curvature is an even function of $\bm k$~\cite{raghu.pra.78.033834.2008}. Thus, when time-reversal and inversion symmetries are broken simultaneously, the total Berry curvature at valleys may be enhanced (gap width increases) or reduced (up to restoration of Dirac point) in comparison with the case, when only one of the symmetries is broken. One can therefore distinguish four different situations, highlighted in Table~\ref{table1}, where the symbol $X$ implies the possibility of the Dirac point restoration. For instance, one can see that changing of only rotation direction or inversion of detuning shifts Dirac points from $\bf{K}$ to ${\bf K}'$ points (or vice versa), while simultaneous change of rotation direction and inversion of detuning will have no effect on the location of the Dirac points.
	We would like to note that the rotation of the waveguide array in Fig.~\ref{fig1}(a) is clockwise $\circlearrowright$,
		and the results correspond to the case \ding{172} in Table~\ref{table1}. 
	
	\begin{figure}[t]
		\centering
		\includegraphics[width=\columnwidth]{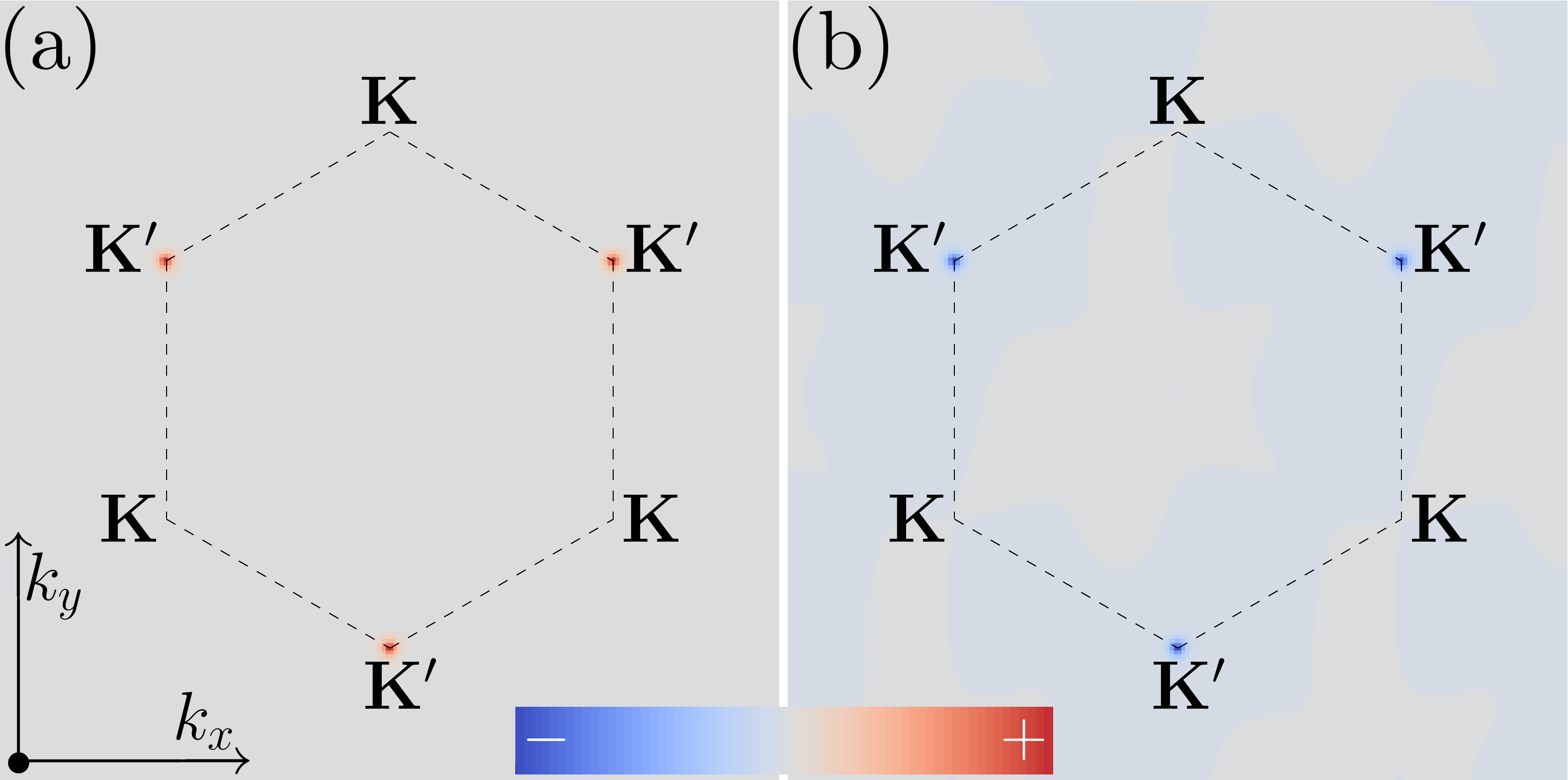}
		\caption{Berry curvature $\mathcal B$ of the bottom band of the lattice with unpaired Dirac cones. Parameters in (a) are: ${a=1.6}$, $Z=6$, $r=0.3$, ${\wp=0.1}$, ${t=0.18}$, and ${{\bm A} (z)= r\Omega[-\cos(\Omega z), \sin(\Omega z)]}$. Parameters in (b) are: $a=1.6$, $Z=6$, $r=0.3$, ${\wp=-0.1}$, ${t=0.18}$, and ${{\bm A} (z)= r\Omega[-\cos(\Omega z), -\sin(\Omega z)]}$. The dashed hexagon represents the first Brillouin zone.}
		\label{fig2}
	\end{figure}
	
	\begin{figure*}[t]
		\centering
		\includegraphics[width=\textwidth]{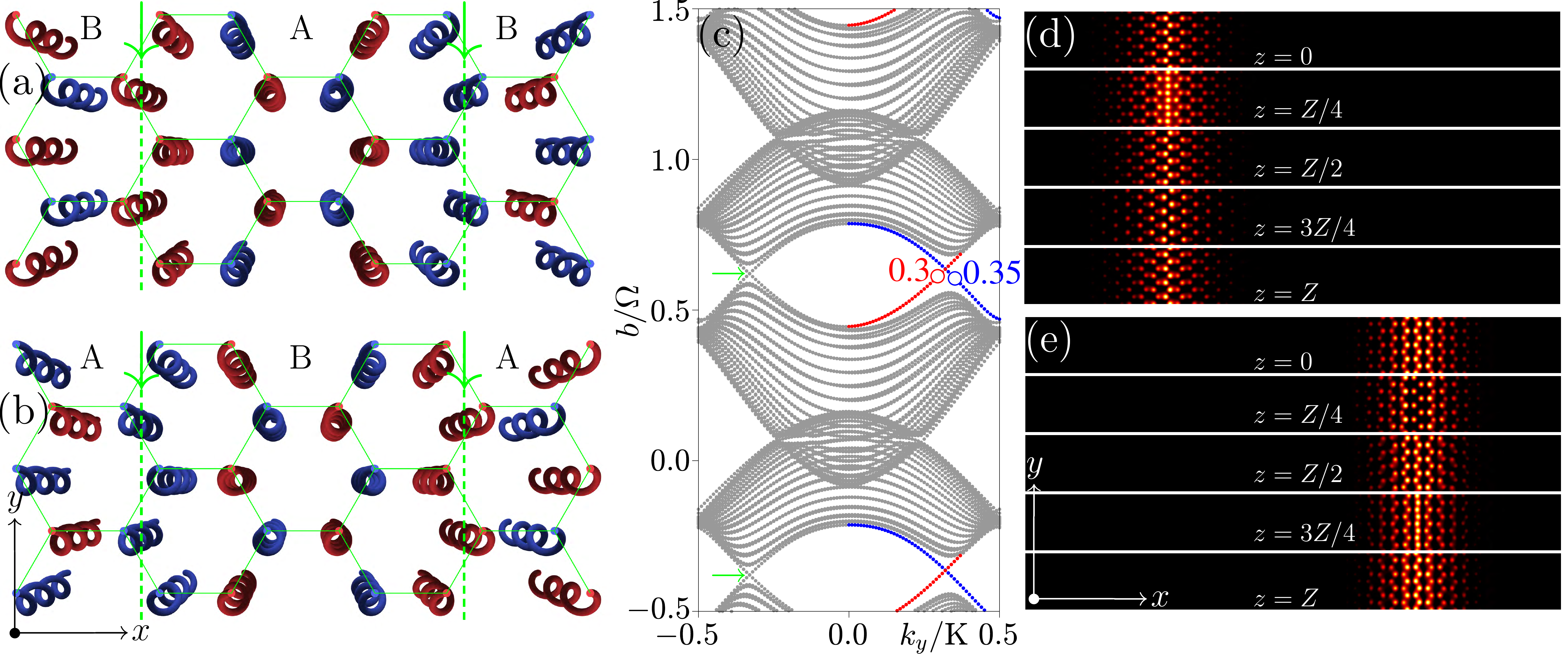}
		\caption{(a) Schematic representation of composite helical waveguide array with two domain walls which are indicated by the vertical dashed green lines. Only waveguides of one type (blue or red) fall onto domain wall. The array is periodic in the $y$ direction. Both detuning and direction of waveguide rotation are opposite in A and B arrays forming the domain wall. (b) A-B-A array configuration with two domain walls. (c) Quasi-propagation constant vs Bloch momentum $k_y$ for B-A-B configuration in (a). Two longitudinal Brillouin zones are shown to stress vertical periodicity of the spectrum. Red and blue dots correspond to the edge states appearing on red and blue domain walls, respectively. Gray dots correspond to bulk modes. (d) Field modulus distribution ${|\psi|}$ of the edge state (three $y$-periods are shown) with ${k_y/{\rm K}=0.3}$ residing on the red domain wall in (c) at different propagation distances within one period $Z$. (e) Edge states with ${k_y/{\rm K}=0.35}$ from blue domain wall at different distances. In all cases $Z=6$, $r_0=0.3$ and $\delta=0.07$.}
		\label{fig3}
	\end{figure*}
	
	\subsection{Topological characterization of the system}
	\label{topo}
	%One can imagine that the band structure will not change if both the helical direction and the detuning change oppositely, i.e., the Dirac cones at $\bf{K}$ points and the valleys at $\bf{K}'$ points are always preserved. However, the Berry curvatures of the valleys are not hold --- it may change from positive to negative or \textit{vice versa}; see cases \ding{172} and \ding{175} in Table~\ref{table1}.
	To characterize topological properties of helical waveguide array with detuning between two sublattices, we utilize tight-binding approximation and the approach summarized in~\cite{zhong.rrp.76.903.2024}. The evolution of the excitations in our system over one $Z$-period (in reciprocal space) can be described by the Floquet evolution operator $U({\bm k}, Z)$ that allows to formally introduce also the effective Hamiltonian $H_{\rm eff}$:
	\begin{align}
		U({\bm k}, Z) & =\mathcal{Z} \exp \left[-i \int_0^Z H({\bm k}, z) d z \right] \notag\\
		&= \exp \left[ -i H_{\rm eff}({\bm k}) Z \right],
	\end{align}
	where $\cal Z$ is the time-ordering operator, and $H({\bm k}, z)$ is the ``instantaneous'' Hamiltonian of the system
	\begin{equation}
		H({\bm k},z) = t
		\begin{bmatrix}
			\wp     & P\\
			P^\dagger  & -\wp
		\end{bmatrix}.
	\end{equation}
	Here 
	\[
	P=\sum_i \exp [-i({\bm A} + {\bm k}) \cdot {\bm e}_i]
	\]
	and
	\[
	{\bm A} (z)= r\Omega[-\cos(\Omega z), \sin(\Omega z)]
	\]
	with ${i=1,2,3}$, 
	${{\bm e}_1= [1,0]a}$, 
	${{\bm e}_2 = [-1/2, -\sqrt{3}/2]a}$, 
	${{\bm e}_3 = [-1/2, \sqrt{3}/2]a}$ are the characteristic vectors of the honeycomb lattice, and ${\bm A} (z)$ is the gauge field arising due to helicity of waveguides, $\wp$ is the on-site ``energy'' detuning~\cite{xiao.prl.99.236809.2007,xue.apr.2.2100013.2021} of two sublattices in tight-binding model, which is different from $\delta$ in Eq.~(\ref{eq2}) but proportional to it, $t$ is the hopping strength. Berry curvature~\cite{xiao.rmp.82.1959.2010} of the band with index $n$ can be introduced as
	\begin{align}
		{\mathcal B}_n(\bm{k})=i \sum_{n' \neq n}  & \left( \frac{\langle u_n|\partial_{k_x} H_{\rm eff}| u_{n'}\rangle \langle u_{n'}|\partial_{k_y} H_{\rm eff}| u_n\rangle}{(b_n-b_{n'})^2} - \right. \notag \\
		&\left. \frac{\langle u_n|\partial_{k_y} H_{\rm eff}| u_{n'}\rangle \langle u_{n'}|\partial_{k_x} H_{\rm eff}| u_n\rangle}{(b_n-b_{n'})^2}\right),
	\end{align}
	where $|u_n \rangle$ and $b$ are the eigenstates and eigenvalues of $H_{\rm eff}$, and $\langle \cdot | \cdot \rangle$ is the inner product.
	Typical distribution $\mathcal{B}_n(\bm{k})$ in the bottom band ${(n=2)}$ is presented in Fig.~\ref{fig2}(a). It is calculated for a set of parameters ${a=1.6}$, ${Z=6}$, ${r=0.3}$, ${\wp=0.1}$, and ${t=0.18}$ at which Dirac cones emerge in $\bf K$ points, while gap is opened between ${\bf K}'$ valleys, where Berry curvature is positive. If one simultaneously changes the sign of detuning $\wp$ and waveguide rotation direction (helicity), the gap again appears between ${\bf K}'$ valleys, but Berry curvature around them becomes negative in the bottom band, as shown in Fig.~\ref{fig2}(b). Topological index characterizing this system is given by the valley Chern number:
	\begin{equation}\label{eq7}
		{\mathcal C}_{v,n}=\frac{1}{2 \pi} \int_v  {\mathcal B}_{v,n}(\bm{k}) d \bm{k},
	\end{equation}
	where $v$ indicates that the integration is carried over close proximity of the valleys. Thus, valley Chern number of each valley in Fig.~\ref{fig2}(a) is $1/2$ while in Fig.~\ref{fig2}(b) it is given by $-1/2$. Therefore, according to the bulk-edge correspondence principle~\cite{lu.np.8.821.2014, ozawa.rmp.91.015006.2019}, if the interface is created between two such arrays in real space, the difference ${1/2-(-1/2)=1}$ of valley Chern numbers predicts the appearance of the topological edge states connecting the upper and bottom bands around the valley in the band structure. Importantly, this situation sharply contrasts with usual valley Hall system, where for nonzero detuning $\wp$ the gap simultaneously opens between all valleys in such a way that Berry curvature has opposite sign around $\bf K$ and ${\bf K}'$ valleys, so that the integral (\ref{eq7}) over the entire Brillouin zone yields zero $\mathcal{C}_n$. Instead, in our case such an integral would be nonzero and opposite in sign for both first and second bands.
	
	\begin{figure*}[ht]
		\centering
		\includegraphics[width=1\textwidth]{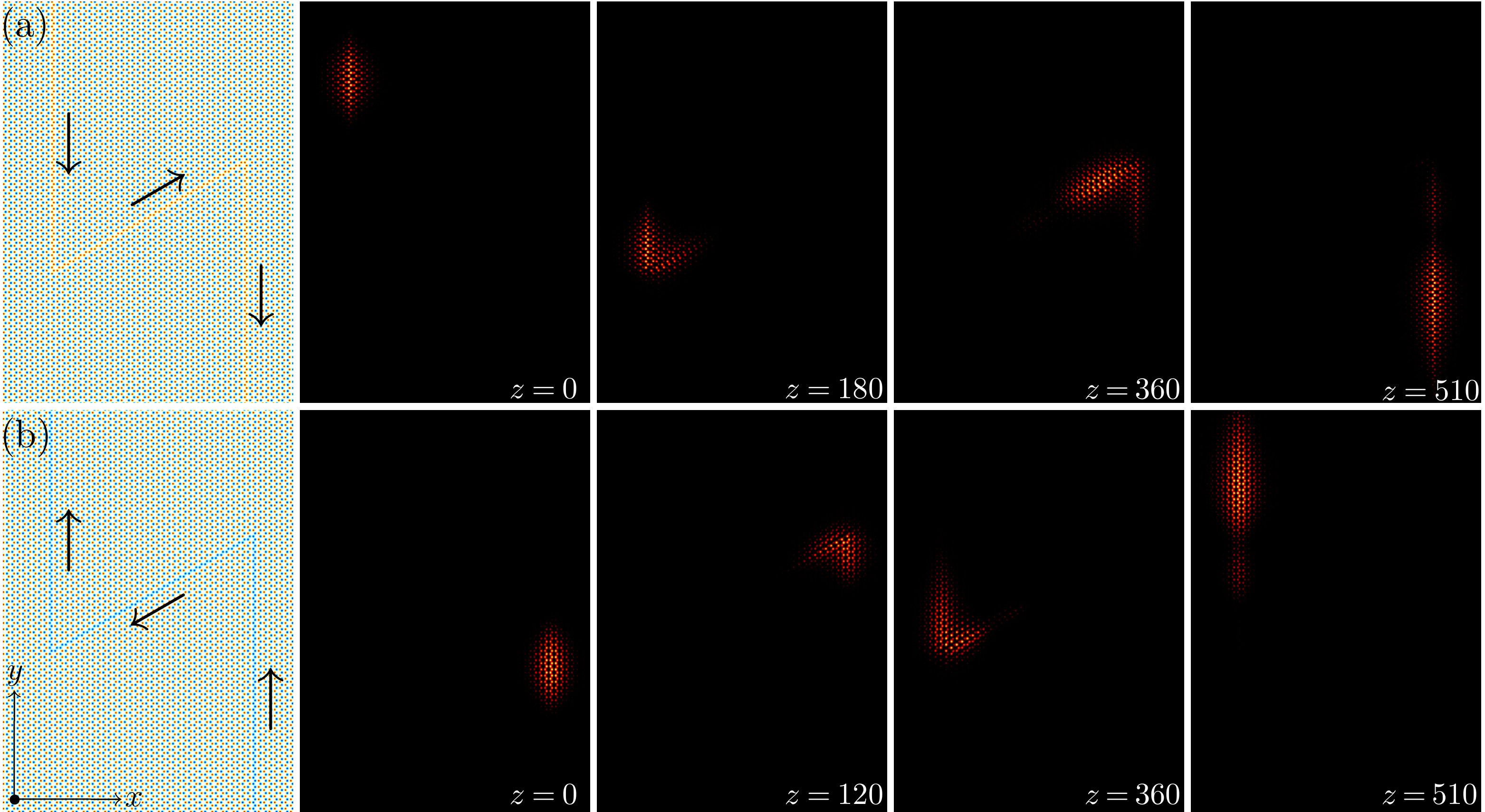}
		\caption{Zigzag-shaped domain walls with sharp corners that are adopted for demonstration of the topological protection of the edge states from the red branch (a) and from the blue branch (b) of Fig.~\ref{fig3}(c). The arrows indicate the direction of motion of the edge states with initial broad envelope at different stages of propagation. Corresponding field modulus distributions at different distances are shown too. The two edge states correspond to ${k_y/{\rm K}=0.3}$ (a) and ${k_y/{\rm K}=0.35}$ (b) respectively.}
		\label{fig4}
	\end{figure*}
	
	\section{Results and discussion}
	
	\subsection{Domain walls and edge states}
	
	To illustrate the possibility of formation of topological edge states in this system with unpaired Dirac points, we constructed the domain wall between helical arrays of types A and B with opposite detuning of two constituent sublattices and with opposite waveguide rotation directions, as illustrated in schematic Figs.~\ref{fig3}(a) and \ref{fig3}(b). As discussed previously, the bands of such arrays with the same index $n$ are characterized by the opposite valley Chern numbers, while Dirac cones in both A and B arrays are located in $\textbf{K}$ points. The arrays can be arranged either into B-A-B [Fig.~\ref{fig3}(a)] or into A-B-A [Fig.~\ref{fig3}(b)] configurations, with waveguides of only one type (red or blue) falling onto domain walls, the latter are highlighted by the green dashed lines. Notice that the waveguides on the domain wall have alternative helix directions since they belong to different arrays A and B. According to the bulk-edge correspondence principle the domain walls between such arrays should support edge states of topological origin. In Fig.~\ref{fig3}(c) we show the dependence of the quasi-propagation constant of all modes in B-A-B configuration on Bloch momentum $k_y$ (along the edge) within first transverse Brillouin zone of width ${{\rm K}=2\pi/\sqrt{3}a}$. Since the array is periodic in $z$, the spectrum is also periodic along the vertical $b$ axis with the period ${\Omega=2\pi/}Z$ (defined by the waveguide rotation frequency). In the band structure in Fig.~\ref{fig3}(c) gray dots correspond to delocalized bulk modes, while red and blue dots correspond to the edge states localized on red and blue domain walls, respectively. Upon calculation of this spectrum we ``glued'' B arrays far away from the domain walls and wide A layer between them by assuming periodic boundary conditions along the $x$-axis. The remarkable feature of the ``projected" spectrum in Fig.~\ref{fig3}(c) is its asymmetry with coexistence of the Dirac cone at ${k_y=-\textrm{K}/3}$ (indicated by the green arrow) and of a gap at ${k_y=+\textrm{K}/3}$, around which unidirectional edge states form that are located at blue and red domain walls, in accordance with line colors. The appearance of such edge states is a remarkable fact, taking into account the absence of the complete gap in the spectrum. We show selected profiles of the edge states from red and blue domain walls in Figs.~\ref{fig3}(d) and \ref{fig3}(e), respectively. Since our Floquet system is $Z$-periodic, such states show exactly periodic evolution on one longitudinal period, with substantial variations of the field modulus distribution in the internal $z$-points. Notice that localization of such edge states is strongest in the center of the ``local'' gap, while when their quasi-propagation constant approaches the band, such states gradually become extended. 
	
	\begin{figure}[h]
		\centering
		\includegraphics[width=\columnwidth]{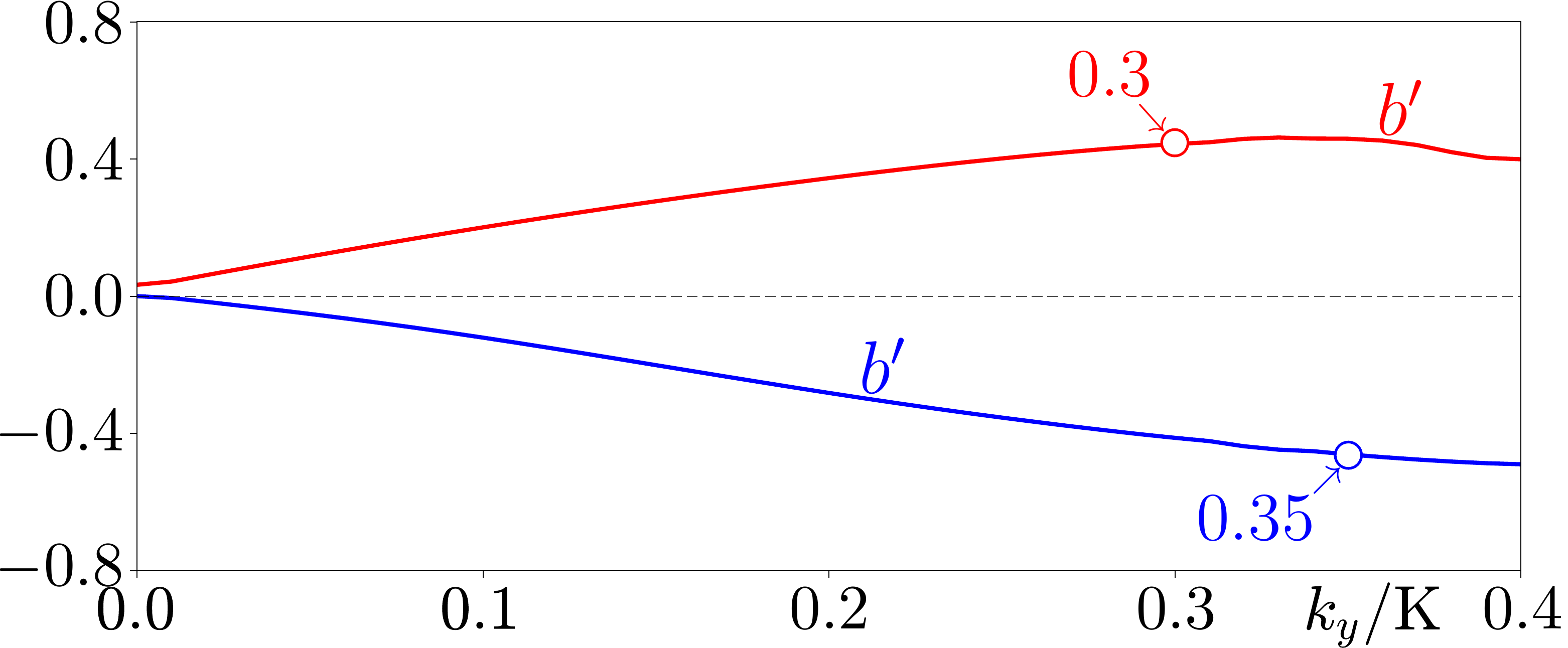}
		\caption{First-order derivative $b'$ of the quasi-propagation constant of the valley Hall edge states at $Z=6$ and $r_0=0.3$.}
		\label{fig5}
	\end{figure}
	
	\begin{figure*}[t]
		\centering
		\includegraphics[width=\textwidth]{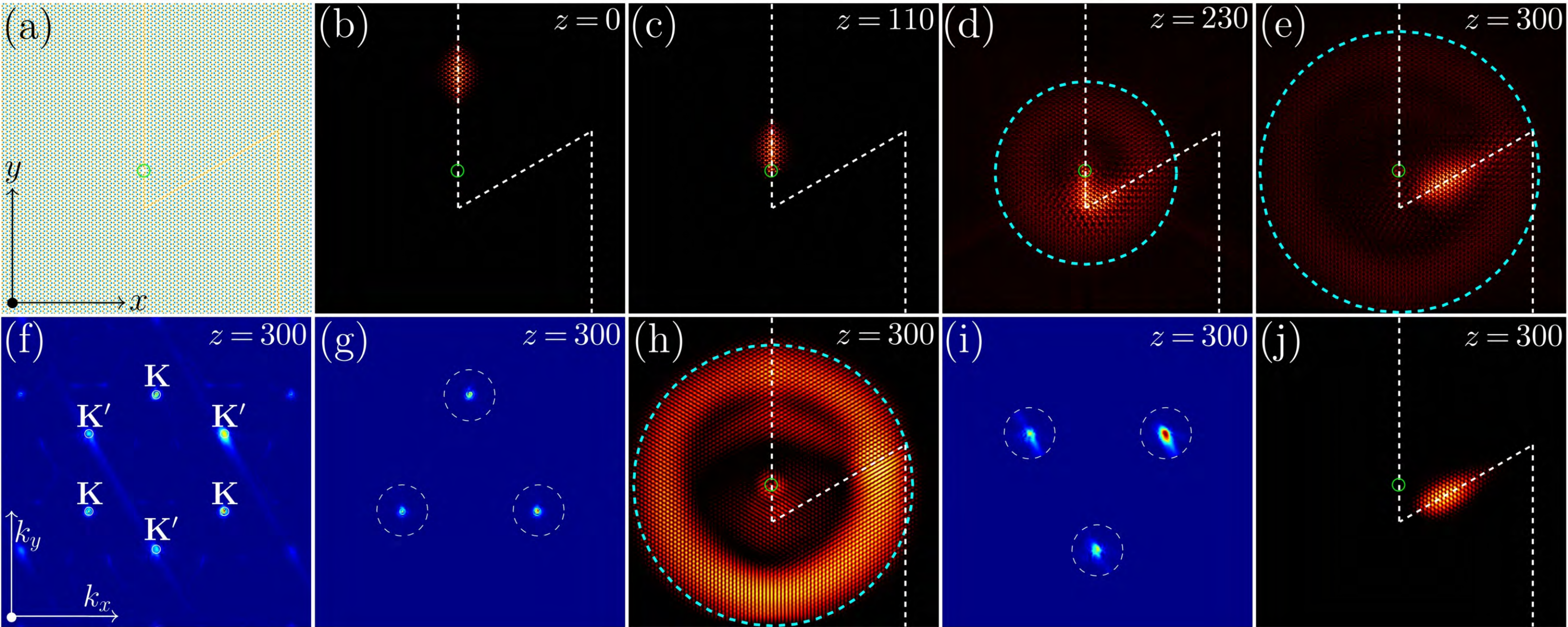}
		\caption{(a) A zigzag domain wall with a defect in the form of missing waveguide indicated by the green circle. (b-e) Spatial field modulus distributions in the edge state at different propagation distances. Cyan dashed circle highlights theoretical radius of the conical diffraction pattern. (f) Total spatial spectrum of the edge state at ${z=300}$. (g) Part of the spectrum at ${z=300}$ where only contribution from regions around $\textbf{K}$ points is left (see white dashed circles), while spectral intensity around ${\bf K}'$ points is set to zero. (h) Field modulus distribution in the form of conical wave produced by spectrum in (g). (i) Part of the spectrum at ${z=300}$ where only contribution from regions around ${\bf K}'$ points is left (see dashed circles), while spectral intensity around ${\bf K}$ points is set to zero. (j) Field modulus distribution of the edge state produced by spectrum in (i). White dashed line represent the zigzag domain wall. The parameters are the same as in Fig.~\ref{fig4}.}
		\label{fig6}
	\end{figure*}
	
	The edge states reported here require for their existence the interface between two lattices with broken inversion symmetry and on this reason they do belong to the class of valley Hall states. However, our structure also combines in a unique fashion the properties typical for valley Hall and Floquet systems. Because time-reversal symmetry in our system is broken due to helicity of waveguides, edge states become truly unidirectional and this allows to avoid the problem typical for usual valley Hall systems that support a pair of counterpropagating edge states that can be coupled by sufficiently narrow edge defects. Our system is free from this limitation and this is reflected in completely different scenario of interaction of edge states with narrow defects that is now accompanied not by reflection of the edge state, but by emission of fraction of power in the form of conical wave (please see Fig. \ref{fig6} below) and passage of considerable fraction of power through the defect. In this sense, our system offers stronger topological protection than conventional valley Hall systems.
	
	\begin{figure*}[t]
		\centering
		\includegraphics[width=1\textwidth]{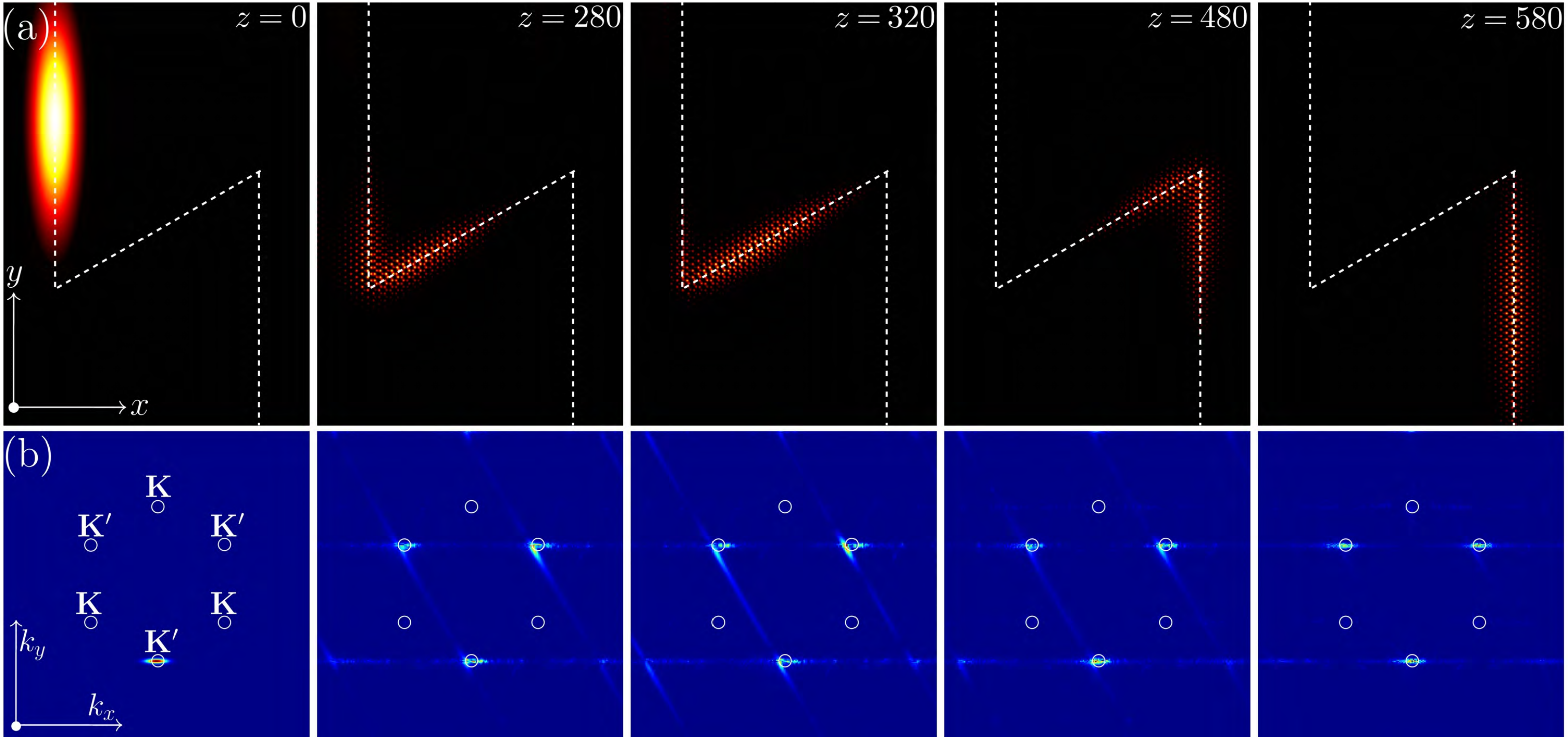}
		\caption{Dynamics of excitation of the valley Hall edge state with a titled elliptical Gaussian beam. (a) shows input field modulus distribution at ${z=0}$ and field modulus distributions at progressively increasing propagation distances, indicating that the excited state bypasses zigzag bend of the domain wall, indicated with white dashed line. (b) shows spectra in momentum space corresponding to spatial distributions in (a). Circles in (b) highlight the valleys in six corners of the first Brillouin zone. Parameters of the array are the same as in Fig.~\ref{fig4}. The input Gaussian beam has width ${w_x=10}$, ${w_y=20\sqrt{5}}$ and initial momentum ${k_x=0}$, ${k_y=-4\pi/3\sqrt{3}a}$.}
		\label{fig7}
	\end{figure*}
	
	\subsection{Topological protection of the edge states}
	
	To confirm the topological nature of the Floquet edge states suggested by associated nonzero valley Chern numbers discussed in Section~\ref{topo}, we designed two types of the domain walls with a zigzag shape that possess two sharp corners (the angle is $60^\circ$). The domain wall in Fig.~\ref{fig4}(a) is similar to red domain wall in Fig.~\ref{fig3}(a), as highlighted by the corresponding line, while the domain wall in Fig.~\ref{fig4}(b) is similar to the blue domain wall in Fig.~\ref{fig3}(a), and it highlighted by the blue line. These domain walls support edge states corresponding to red and blue branches of Fig.~\ref{fig3}(c) that have opposite group velocities ${v_g=-b'=-db/dk_y}$. Since ${b'>0}$ for the red edge state (see red curve in Fig.~\ref{fig5}), it moves in the negative $y$-direction, while its blue counterpart with ${b'<0}$ moves in the positive direction of $y$-axis (see blue curve in Fig.~\ref{fig5}). The propagation dynamics of corresponding edge states with initial broad envelopes is illustrated in Figs.~\ref{fig4}(a) and \ref{fig4}(b), the direction of motion of corresponding states is highlighted by the black arrows on schematics with zigzag domain walls. Field modulus distributions at different distances clearly show that both red and blue edge states can circumvent sharp corners during propagation without any noticeable backscattering, that illustrates their topological protection, at least for such type of edge deformations. Notice that small tails visible in field modulus distributions at ${z=510}$ are due to dispersion of the wavepacket, as they move in the same direction as wavepacket and with the same velocity. This is intriguing result, taking into account the absence of complete gap in the spectrum of the system. We would like to note that in dynamics illustrated in Fig. \ref{fig4} the edge state shows neither inter-valley scattering nor it excites bulk states because this state is taken quite close to the $\textbf{K}'$ point in momentum space, and its quasi-propagation constant is located practically in the center of the gap. However, if the input edge state is taken further from the $\textbf{K}'$ point, for example with ${k_y=0.2 \rm K}$, such that its quasi-propagation constant can overlap with quasi-propagation constants of the bulk modes, a fraction of power can be radiated into the bulk when the state bypasses the sharp corner (see \textbf{Supplemental Information} for corresponding results).
	
	It is a well-established fact that in valley Hall systems the inter-valley scattering can be induced by strongly localized defects, while broad slowly changing defects cannot induce such a scattering. To test the behaviour of the edge states in our system with respect to strongly localized perturbations, we removed one waveguide located on the domain wall, as illustrated by the green circle in Fig.~\ref{fig6}. The incident edge state at ${z=0}$ in Fig.~\ref{fig6}(b) is the same as in Fig.~\ref{fig4}(a) and it encounters the defect around ${z=110}$ [Fig.~\ref{fig6}(c)] while propagating along the domain wall. The interaction between the defect and the edge state shows that the edge state cannot bypass the defect without radiation [see Figs.~\ref{fig6}(d) and \ref{fig6}(e)]. Remarkably, because edge states in our system are unidirectional, no backward reflection at the defect occurs. Instead, inter-valley scattering leads to the excitation of waves in close proximity of the Dirac cone at the $\bf K$ point, and further propagation of such waves is accompanied by their conical diffraction resulting in gradually expanding ring visible in field modulus distributions at ${z=230}$ and ${z=300}$. The dashed cyan circle in Figs.~\ref{fig6}(d) and \ref{fig6}(e) [and also in Fig.~\ref{fig6}(h)] illustrates theoretically evaluated radius of the conical diffraction pattern ${\sim |b'_{\bf K} (z-110)|}$ with ${b'_{\bf K}\sim 0.45}$ being the first-order derivative of the quasi-propagation constant around Dirac cone at the $\bf K$ point. Remarkably, only a fraction of input power is emitted in the form of conical wave after interaction width defect, while the rest of the power is organized into edge state that bypassed the defect and keeps moving along the domain wall - a consequence of the unidirectional nature of the edge states in our system. To illustrate this, we take the exemplary field distribution at $z=300$ and display its total spatial spectrum in Fig.~\ref{fig6}(f). Clearly, there is light around both ${\bf K}$ and ${\bf K}'$ points. If we isolate contributions from ${\bf K}$ points by applying the filter of the form
	\begin{equation}
		F_{\bf K}({\bm k})=\left\{
		\begin{split}
			&1, ~|{\bm k} -{\bf K}|\le r_k,\\
			&0, ~{\rm other~places},
		\end{split}
		\right.
	\end{equation}
	to total spectrum, where $r_k=4\sqrt{3}\pi/27a$ is the radius of the circular filter, we will obtain the spectrum depicted in Fig.~\ref{fig6}(g) (white dashed circles show the regions outside which the spectral intensity was set to zero), that corresponds to field distribution shown in Fig.~\ref{fig6}(h) in the form of clear conical diffraction pattern. In contrast, if the filter $F_{\bf K'}({\bm k})$ is applied that isolates contributions from ${\bf K}'$ points, see the resulting spectrum in Fig.~\ref{fig6}(i), in spatial domain one obtains spatial distribution depicted in Fig.~\ref{fig6}(j) that clearly shows that edge state partially bypassed the defect and keeps propagating along the domain wall after emission of a fraction of its initial power in the form of conical wave. Because the edge states reported here bypass strongly localized defects with emission of fraction of power in the form of conical waves, but at the same time they can still clearly bypass sharp corners practically without radiation, akin to edge states in valley Hall systems reported previously~\cite{tang.oe.29.39755.2021, ren.nano.10.3559.2021}, we attribute them to this last class of edge states.
	
	\subsection{Excitation of the valley Hall edge states}
	
	We would like to note that the valley Hall edge states reported here can be efficiently excited by a Gaussian beam with proper initial tilt (i.e. we multiply the input field distribution with $e^{ik_xx+ik_yy}$ term). In Fig.~\ref{fig7}, we display an example of such excitation dynamics. The elliptical Gaussian beam is launched at the left branch of the zigzag domain wall [see field modulus distribution in Fig.~\ref{fig7}(a) and corresponding spatial spectrum in Fig.~\ref{fig7}(b)]. This input beam quickly reshapes and after emission of some radiation efficiently excites moving valley Hall edge state with Bloch momentum $k_y$ corresponding to the initial momentum of the Gaussian beam. Excited moving edge state bypasses zigzag bend of the domain wall without reflections. The spectra at different distances presented in Fig.~\ref{fig7}(b) demonstrate that the beam is always concentrated around the ${\bf K}'$ points. The efficiency of this excitation process strongly depends on the momentum of initial Gaussian beam. For example, if its spectrum of the input beam is shifted away from ${\bf K}'$ valleys, one observes considerable diffractive broadening in real space without obvious excitation of localized edge states. For instance, if the input beam excites the ${\bf K}$ valley for a given type of the domain wall, one observes conical diffraction in spatial domain \cite{peleg.prl.98.103901.2007}.
	
	\section{Conclusion and outlook}
	
	Summarizing, we have shown that helical waveguide arrays with detuning between two sublattices, where time-reversal and inversion symmetries are simultaneously broken, can feature unpaired Dirac points in their linear spectra. The formation of the unpaired Dirac points in $\bf K$ or ${\bf K}'$ points of the Brillouin zone can be controlled by the sign of detuning and waveguide helicity. Even though the location of Dirac cones does not change upon simultaneous change of the sign of detuning and waveguide helicity, the sign of Berry curvature changes that allows to construct the interface between two arrays with opposite valley Chern numbers, hosting unidirectional topological edge states. These states are topologically protected despite the absence of the complete gap in the spectrum of the system. These results suggest a platform and pave the way to exploration of various nonlinear phenomena in Floquet insulators with unpaired Dirac points. For instance, we would like to point out that in the past decade a variety of hybrid topological edge solitons have been reported in topological systems based on helical waveguide arrays~\cite{ablowitz.pra.96.043868.2017, lumer.prl.111.243905.2013, ivanov.acs.7.735.2020, ivanov.ol.45.1459.2020, ivanov.ol.45.2271.2020, ivanov.pra.103.053507.2021, ivanov.lpr.202100398, mukherjee.science.368.856.2020}, in valley Hall arrays with straight waveguides~\cite{smirnova.lpr.13.1900223.2019, zhong.ap.3.056001.2021, ren.nano.10.3559.2021, tang.oe.29.39755.2021, tang.chaos.161.112364.2022, tang.rrp.74.504.2022, tian.fop.17.53503.2022} and other systems~\cite{kartashov.optica.3.1228.2016, gulevich.sr.7.1780.2017, li.prb.97.081103.2018, zhang.pra.99.053836.2019, zhang.prl.123.254103.2019}, but all these structures are characterized by the presence of complete topological gaps and, as a rule, are developed on lattices possessing paired Dirac points. Similarly, generation of higher harmonics and topological lasers may acquire very unusual features in systems with unpaired Dirac points. 
	Last but not least, the unpaired Dirac cones ensure that the inter-valley scattering will not happen in the system developed in the work which may bring a different view on the conclusions reported very recently~\cite{rosiek.np.17.386.2023,rechtsman.np.17.383.2023}.
	We believe that the system with unpaired Dirac cones may provide a novel platform for investigating higher-order topological states~\cite{xie.nrp.3.520.2021,lin.nrp.5.483.2023,zhang.elight.3.5.2023} and may help to observe the unpaired Weyl points~\cite{guo.elight.3.2.2023}.
	
	\begin{backmatter}
		\bmsection{Funding} This work was supported by the Natural Science Basic Research Program of Shaanxi Province (2024JC-JCQN-06), the National Natural Science Foundation of China (12074308, 12304370), and the Fundamental Research Funds for the Central Universities (sxzy012024146). The work of Y.V.K. was supported by the research project FFUU-2024-0003 of the Institute of Spectroscopy of the Russian Academy of Sciences.
		
		\bmsection{Acknowledgments} The authors appreciate the three anonymous reviewers for their insightful and helpful comments which improve the work greatly.
		
		\bmsection{Disclosures} The authors declare no conflicts of interest.
		
		\bmsection{Data availability} Data underlying the results presented in this paper are not publicly available at this time but may be obtained from the authors upon reasonable request.
		
	\end{backmatter}

%	% Bibliography
%	\bibliography{my_library}

\end{document}